\documentclass[%
 aip,
 amsmath,amssymb,
 reprint,%
]{revtex4-1}

\usepackage{graphicx}
\usepackage{dcolumn}
\usepackage{bm}

\usepackage[utf8]{inputenc}
\usepackage[T1]{fontenc}

\usepackage{mathptmx}
\usepackage{siunitx}
\usepackage{graphicx}
\usepackage{dcolumn}
\usepackage{braket}
\usepackage{xfrac}
\usepackage{hyperref}
\usepackage[usenames,dvipsnames]{color}
\usepackage{tikz-cd}
\usepackage{float}
\usepackage[nameinlink,capitalise]{cleveref}
\usepackage{mathtools}
\usepackage{amsmath}
\usepackage{bm}
\DeclareSIUnit\gauss{G}
\DeclareSIUnit{\au}{{a.u.}}
\hypersetup{
colorlinks=true,
linkcolor=blue,
citecolor=blue,
filecolor=green,
urlcolor=blue,
}

\begin{document}


\title{Stereodynamics of rotationally inelastic scattering in cold He+HD collisions}

\author{Masato Morita}

\author{Naduvalath Balakrishnan}%
\email{naduvala@unlv.nevada.edu}
\affiliation{ 
Department of Chemistry and Biochemistry, University of Nevada, Las Vegas, Nevada 89154, USA
}%


\begin{abstract}

Stereodynamics of cold collisions has become a fertile ground for quantized studies of molecular collisions and control of the collision outcome. 
A benchmark process for stereodynamic control is rotational transition in He+HD collisions. This process was recently probed experimentally by Perreault {\it et al}. by examining quenching from $j=2$ to $j'=0$ state in the $v=1$ vibrational manifold. 
Here, through explicit quantum scattering calculations on a highly accurate {\it ab initio}  interaction potential for  He+H$_2$, 
we reveal how a combination of two shape resonances arising from $l=1$ and $l=2$ partial waves controls the stereodynamic outcome rather than a single $l=2$ partial wave attributed in the experiment. 
Further, for collision energies below 0.5 cm$^{-1}$, it is shown that  stereodynamic preference for integral cross section follows a simple universal trend. 
 
\end{abstract}

\maketitle

Molecular hydrogen is the most abundant molecule in the universe and it is also the simplest neutral molecule. 
Collisions of H$_2$ and its isotopomers with He and H$_2$ have long served as prototypes for inelastic rotational and vibrational 
transitions in molecules while H+H$_2$ and its isotopic counterparts have served as benchmarks for reactive scattering. Strong interests in these systems continue as they are amenable to precise calculations of their interaction potential and highly resolved experimental studies. Indeed, the HD molecule has been the topic of a number of recent studies focusing on the geometric phase effect in chemical reactions~\cite{Kendrick_PhysRevLett.115.153201, Yuan1289,Xie767} and stereodynamics of cold molecular collisions.~\cite{2017_Science_Perreault,2018_NatChem_Perreault,SARP_HD-He,err_SARP_HD-He,2018_PRL_Croft,2019_JCP_Croft,2019_PRL_Jambrina} The latter topic has attracted considerable attention as recent developments in intra-beam technologies~\cite{doi:10.1021/acs.jpclett.7b02378,doi:10.1063/5.0007382} allow molecular collisions to be explored near 1 K. In this regime collisions of light molecules like HD occur through a few low angular momentum partial waves. Stereodynamics of such collisions is controlled by isolated resonances and allow theoretical predictions to be benchmarked against experiments.~\cite{2018_PRL_Croft,2019_JCP_Croft,2019_PRL_Jambrina} Experiments on cold and ultracold molecules~\cite{Rev_Carr08,Rev_Krems08,Rev_Ye2014,Rev_Bala16,Rev_Bohn17} have largely focused on dimers of alkali metal atoms prepared using photoassociation or Feshbach resonance methods and HD presents an excellent test case for benchmarking theory and experiment 
for a prototype non-alkali metal dimer system.

Recently the stereodynamics of rotationally inelastic scattering of HD  with He was examined by Perreault {\it et al}.~\cite{SARP_HD-He,err_SARP_HD-He} in which the Stark induced adiabatic Raman Passage (SARP) technique was used to prepare HD initially  in the $v=1,\ j=2$ ro-vibrational state with control of its alignment and orientation. 
By analyzing the angular  distribution of the inelastically
scattered HD in the $v'=1, j'=0$ state they concluded that only even partial waves contribute, and in particular, an isolated incoming $d$-wave ($l=2$) shape resonance near a collision energy of 1-3 cm$^{-1}$ accounts for the observed angular distribution. 
However, the experiment does not provide energy resolved measurements although
90\% of the collisions were attributed to energies $<5$ K,
drastically limiting the number of partial waves.

In this communication we provide a theoretical analysis of stereodynamics of rotationally inelastic scattering of HD 
($v=1, j=2  \to v'=1, j'=0$)
by collisions with He 
based on  rigorous quantum scattering calculations on a highly accurate {\it ab initio} potential energy surface (PES) for the HeH$_2$ complex.
It is found that the integral cross section (ICS) for $j=2 \to j'=0$ transition exhibits a 
stereodynamic preference with a universal trend below $<0.8$ cm$^{-1}$ that also features a pronounced peak due to an incoming $p$-wave ($l=1$) shape resonance. 
In the energy range of 1-3 cm$^{-1}$ the ICS displays a weak shoulder feature arising from a 
$l=2$ shape resonance. The stereodynamic effect in this regime and at higher energies is significantly influenced by the interference between different partial waves. We show that dominant features of the experimental angular distribution  measured by Pereault {\it et al}.~\cite{SARP_HD-He,err_SARP_HD-He}  include signatures of both $l=1$ (odd) and $l=2$ (even) partial waves shape resonances.

Rotationally inelastic scattering of HD by He has been extensively reported in the literature primarily for astrophysical applications.~\cite{Flower1999,Rueff2000,Nolte_2011} The HD molecule breaks the homonuclear symmetry of H$_2$  and angular dependence of the interaction potential for He+HD features both  even and odd terms compared to only even terms for He+H$_2$. 
In this study, we adopt a modified version of the PES for the HeH$_2$ system developed by Bakr, Smith and Patkowski~\cite{3D_PES} referred to as the BSP3 PES~\cite{BSP3}. 
Its uncertainty in the van der Walls well region is estimated to be less than 0.04 K, well below the mean collision energies involved in the experiments of Perreault {\it et al}.~\cite{SARP_HD-He}
The quantum scattering calculations are carried out based on the close-coupling (CC) scheme using the MOLSCAT code~\cite{MOLSCAT} (see also supplementary material).

In the SARP experiment~\cite{SARP_HD-He} the HD molecule is prepared in a rotational state $j$ with a specific projection component $m$ as $|j,m>$ or a coherent superposition of rotational states with different $m$ \cite{2017_Science_Perreault,2019_JCP_Croft},
where the z-axis for the projection of $j$ is parallel to the initial relative velocity for the collision.
For $j=2$, in the experimental H-SARP preparation (HD bond is preferentially aligned parallel to the initial relative velocity),  the rotational state is described as $|j, m>=|2, 0>$  
while for the V-SARP preparation (HD bond is preferentially aligned perpendicular to the initial relative velocity) the rotational state is described as $\sqrt{3/8}(|2,-2>+|2, 2>)-(1/2)|2, 0>$.  
The HD is also prepared in a state $\sqrt{1/2}(|2,-1>+|2, 1>)$ using X-SARP in the experiment.~\cite{SARP_HD-He}

For collisions between a structureless atom ($^1$S) and a closed-shell molecule ($^1\Sigma$) the scattering amplitude for rotational transition between oriented ($m$-specified) rotational states is given by,~\cite{1952_Blatt_Biedenharn_Rev,1960_Dalgarno_Arthurs,1977_Alexander_1,1977_Alexander_2} 
\begin{equation}
 \begin{split}
&f_{{j m}\to{j' m'}} (\theta,\phi,E) \\
& = \sqrt{\pi}(-1)^{j+j'}   \sum_{J=0}  \sum_{l=|J-j|}^{J+j}   \sum_{l'=|J-j'|}^{J+j'} 
 i^{l'-l}(2J+1)\sqrt{2l+1} \\
& \times \begin{pmatrix} j & l & J\\ m & 0 & -m \end{pmatrix}
\begin{pmatrix} j' & l' & J\\ m' & m-m' & -m \end{pmatrix}
 T^J_{jl,j'l'}(E) Y_{l'm-m'}(\theta,\phi),
\end{split}
\label{eq:Amp}
\end{equation}
where $\theta$ and $\phi$ are the scattering polar angles in the center of mass frame, $E$ is the total energy, $J$ is the quantum number of the total angular momentum of the collision complex defined as $\bm{J}={\bm j}+{\bm l}$ with ${\bm l}$ denoting the orbital angular momentum between collision partners, 
$T(E)$ is the T-matrix obtained from the S-matrix as $T(E)=1-S(E)$,
and $Y_{l'\,m-m'}$ denote spherical harmonics.

To obtain differential and integral cross sections, DCS and ICS, it is necessary to evaluate the square of modulus of the scattering amplitude. The DCS is given as 
\begin{equation}
\frac{d\sigma_{{j m}\to{j' m'}}}{d\Omega}=
\frac{| f_{{j m}\to{j' m'}} (\theta,\phi,E)|^2}{k^2},
\label{eq:DCSfull}
\end{equation}
where $k$ is the magnitude of the initial wave vector. Since the $\phi$-dependence is not observed in the experiment, our interest is the $\theta$-dependence obtained after averaging over $\phi$:
\begin{align}
\frac{d\sigma_{{j m}\to{j' m'}}}{d\theta} &= \text{sin}\theta \int_0^{2\pi} \frac{d\sigma_{{j m}\to{j' m'}}}{d\Omega} d\phi  \nonumber \\
&= 2\pi \text{sin}\theta \frac{|f_{{j m}\to{j' m'}}(\theta,E)|^2}{k^2},
\label{eq:DCS}
\end{align}
where $f_{{j m}\to{j' m'}}(\theta,E)$ is the $\theta$-dependent scattering amplitude obtained by excluding the $\phi$-dependent phase factor $\text{exp}\{-i(m-m')\phi\}$ from $f_{{j m}\to{j' m'}} (\theta,\phi,E)$ defined in \cref{eq:Amp}.   
The ICS ($\sigma_{{j m}\to{j' m'}}$) is obtained by an integral of the DCS over $\theta$ from 0 to $\pi$. 

\begin{figure}[b!]
\begin{center}
\includegraphics[scale=0.42]{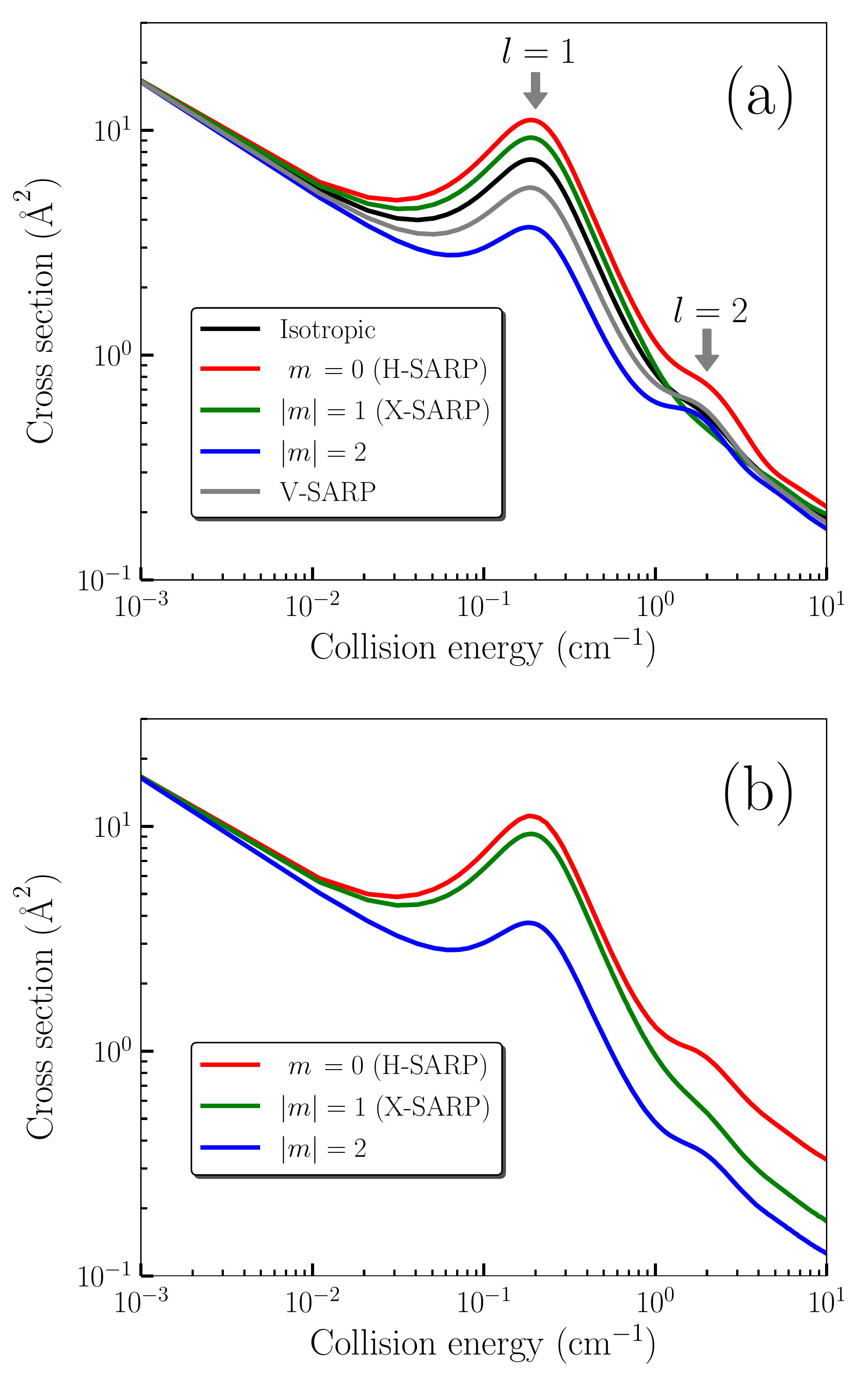}
\end{center}
\caption{State-to-state integral cross section for HD ($v=1,j=2 \to v'=1,j'=0$) by collisions with He. 
(a) Initial HD orientation ($m$) dependence is displayed with red ($m=0$), green ($|m|=1$), and blue ($|m|=2$). 
Results for V-SARP preparation and isotropic collisions (non-polarized) are displayed in gray and black curves, respectively. 
(b) Initial HD orientation ($m$) dependence ignoring the interference of different terms in the scattering amplitude (\cref{eq:Amp}).
}
\label{fig:ICS}
\end{figure}

Our goal is to examine the quenching of HD from $j=2 \to j'=0$ ($m'=0$) prepared in various initial alignments and orientations. Thus the $m$-dependence ($m=0,\ \pm1,\ \text{and}\ \pm2$) of $f_{{j=2, m}\to{j'=0, m'=0}}$ and associated 
cross sections are fundamental quantities because the cross section for any initial orientation or alignment is given by the sum of these fundamental cross sections. 
For example, the ICSs with the V-SARP and X-SARP preparations are given using the square of modulus of the coherent expansion coefficients for the initial rotational states as
$\sigma^\text{V} = (3/8)(\sigma_{{2, -2}\to{0, 0}}+\sigma_{{2, 2}\to{0, 0}}) +(1/4)\sigma_{{2, 0}\to{0, 0}}$ \\
and $\sigma^\text{X}=(1/2)(\sigma_{{2, -1}\to{0, 0}}+\sigma_{{2, 1}\to{0, 0}})$, respectively. Note that $\sigma^\text{X}=\sigma_{{2, -1}\to{0, 0}}=\sigma_{{2, 1}\to{0, 0}}$. 

\Cref{fig:ICS} (a) shows the ICS for $j=2 \to j'=0$ as a function of the collision energy for non-polarized (isotropic)
collisions and different initial orientations ($|m|=0, 1, \text{and}\ 2$) as well as the V-SARP preparation. Two features 
are notable:  a prominent peak near 0.2 cm$^{-1}$  from an $l=1$ partial wave and a shoulder at 1-3 cm$^{-1}$ from an $l=2$ partial wave. 
For $j=2$, the partial wave $l=0$ occurs for $J=2$, the $l=1$ contribution arises from $J=1$ and 3, and $l=2$ from $J=0$, 2 and 4.  
For both $l=1$ and 2, the component associated with the highest $J$ and outgoing partial wave $l'$ given as $l+2$ has the largest contribution (see Fig.~S1 in supplementary material). We note that $J=l'$ due to $j'=0$ and the conservation of $J$.

Previous calculations by  Zhou {\it et al}.~\cite{ICS_HD-He_Boyi}  
for ($j=2 \to j'=0$ in $v=0$)
using the BSP PES of Bakr {\it et al}.~\cite{3D_PES} 
have yielded similar ICS as our isotropic result, including the pronounced ($l=1$) and the weak shoulder ($l=2$) resonances.
Additional tests by scaling the BSP3 PES by $\pm 1\%$ did not alter the resonance structures.
We tested three other available HeH$_2$ PESs~\cite{3D_PES,HD_PEC_modify,HeH2_Muchnick_Russek} and all yielded very similar results as the BSP3 PES employed in this study. For these tests we used the modified version of the PES by Boothroyd {\it et al}.~\cite{HD_PEC_modify} and Muchnick and Russek~\cite{HeH2_Muchnick_Russek} reported by Bakr {\it et al}.~\cite{3D_PES}
A comparison of the ICSs from these PESs indicates that the resonance features are robust to possible uncertainties in the PES (see Fig.~S2  in supplementary material). 
Moreover, recent line-shape parameters of H$_2$ and HD immersed in He evaluated using BSP3 PES achieved excellent (subpercent) agreement with ultra-accurate  experimental results~\cite{BSP3_2020_PRA,BSP3_2020_MA} validating the spectroscopic-level accuracy of the PES used in this study.

Around the primary peak ($l=1$), the ICS clearly shows a stereodynamic preference ($m$-dependence). The ICS with $m=0$ is the largest in the whole energy range while $|m|=2$ is the lowest except around round 2 cm$^{-1}$ 
(in the vicinity of the $l=2$ resonance)
where the order of  $|m|=1$ and 2 is reversed. 
As 75 \% of the V-SARP ICS comes from  $m=\pm2$, the H-SARP ICS ($m=0$) should be larger in the entire energy range, consistent with observed experimental results.
In the lowest energy region, the stereodynamic preference disappears as $s$-wave scattering dominates 
and the Wigner threshold behavior of inelastic cross section emerges (see also Fig.~S1 in supplementary material).

Now we analyze the origin of the stereodynamic preference for the $j=2,m \to j'=0, m'=0$  transition. 
In the low energy regime, the scattering amplitude (\cref{eq:Amp}) is dominated by $s$-wave ($l=0$,\,$J=l'=2$) in the incident channel. The initial orientation ($m$) dependence of the scattering amplitude comes from the two 3-j symbols. Note that $m$-dependence in the spherical harmonic $Y_{l'm}$ will not remain in the ICS because the integral of $|Y_{l'm}|^2$ over $\theta$ yields  unity in evaluating ICS.
The product of the two 3-j symbols results in the $m$-independent factor of $1/5$, while each 3-j symbol has a $m$-dependent phase factor~\cite{Zare_Angular} (see also supplementary material). 
A striking feature is that the initial orientation independence comes from purely algebraic character of the scattering amplitude, thus the $m$-independence of the ICS ($j=2 \to j'=0$) is universal for 
atom ($^1$S) + molecule ($^1\Sigma$)
systems  in the low energy limit. Also this $m$-independence for $l=0$  holds at any collision energy.

From the above analysis, it is clear that the origin of the stereodynamic preference observed at low energies (below 0.8 cm$^{-1}$) in \Cref{fig:ICS} (a) is  due to the $l=1$ partial wave. 
For $j=2$ there are two terms associated with $l=1$ in the scattering amplitude: $J=l'=1$ and $J=l'=3$. Again $m$-dependence for these terms comes from the two 3j-symbols. For $J=l'=1$, the product of the two 3j-symbols is $m$-dependent as $-\sqrt{(4-m^2)/90}$ with $m=0$ and $|m|=1$, and zero with $|m|=2$.~\cite{Zare_Angular, Mathematica}  As shown in \cref{eq:DCS}, the square of the modulus of the scattering amplitude determines the magnitude of the cross section, thus the ratio of the ICS is 4:3 for $m=0$ vs $|m|=1$. Similarly, for  $J=l'=3$, the ratio of the ICSs becomes 9:8:5 for $m=0$, $|m|=1$ and $|m|=2$, respectively~\cite{Mathematica} (see also supplementary material).  Importantly, the stereodynamic preference for $m=0$ and $|m|=1$ is similar for both $J=1$ and 3. We note that, for the ICS, there is no interference effect from the cross
terms arising from these two terms and that with $l=0$ ($J=2$). 


Thus, a universal trend exists for the stereodynamic preference for the ICS in low energy  collisional rotational quenching ($j=2 \to j'=0$). 
We observe no $m$-dependence in the $s$-wave regime followed by a gradual universal trend in $m$-dependence with increasing collision energy due to the contribution from an $l=1$ partial wave ($m=0$ (H-SARP) $>$  $|m|=1$ (X-SARP)  $>$ (V-SARP) $>$ $|m|=2$). This universal trend is regardless of the presence of an $l=1$ shape resonance but in the present case, the shape resonance accentuates this feature. The $m$-resolved ICS ignoring cross terms in the square of modulus of the scattering amplitude is shown in \Cref{fig:ICS} (b).  In comparison with (a),
we observe that  it is possible to describe the ICS accurately without the interference terms (cross terms) of the scattering amplitude until the region of the $l=1$ shape resonance. At higher energies (above 0.8 cm$^{-1}$ in the present case),
the interference effect between higher-order partial waves plays a crucial role in the stereodynamics.


As stated above, the experimental study \cite{SARP_HD-He} concluded that only even partial waves, in particular, a resonantly enhanced $l=2$ partial wave, contributes to the 
inelastically scattered HD angular distribution. In contrast, the calculated ICS reveals contribution from the $l=1$ resonance 
is significant in the relevant collision energy regime and its background contribution is non-negligible even in the region of the $l=2$ resonance, $\sim$ 1-5 cm$^{-1}$ (see Fig.~S1 in supplementary material).
However, if the intensities of the $l=1$ and  $l=2$ resonances are reversed  or if collisions below 1 cm$^{-1}$ are ineffective in causing rotational relaxation under the experimental conditions then it is conceivable that experimental results will show primarily  signatures of the $l=2$ resonance. As discussed earlier, resonance features are very stable and robust across all available HeH$_2$ PESs.

\begin{figure}[b!]
\begin{center}
\includegraphics[scale=0.4]{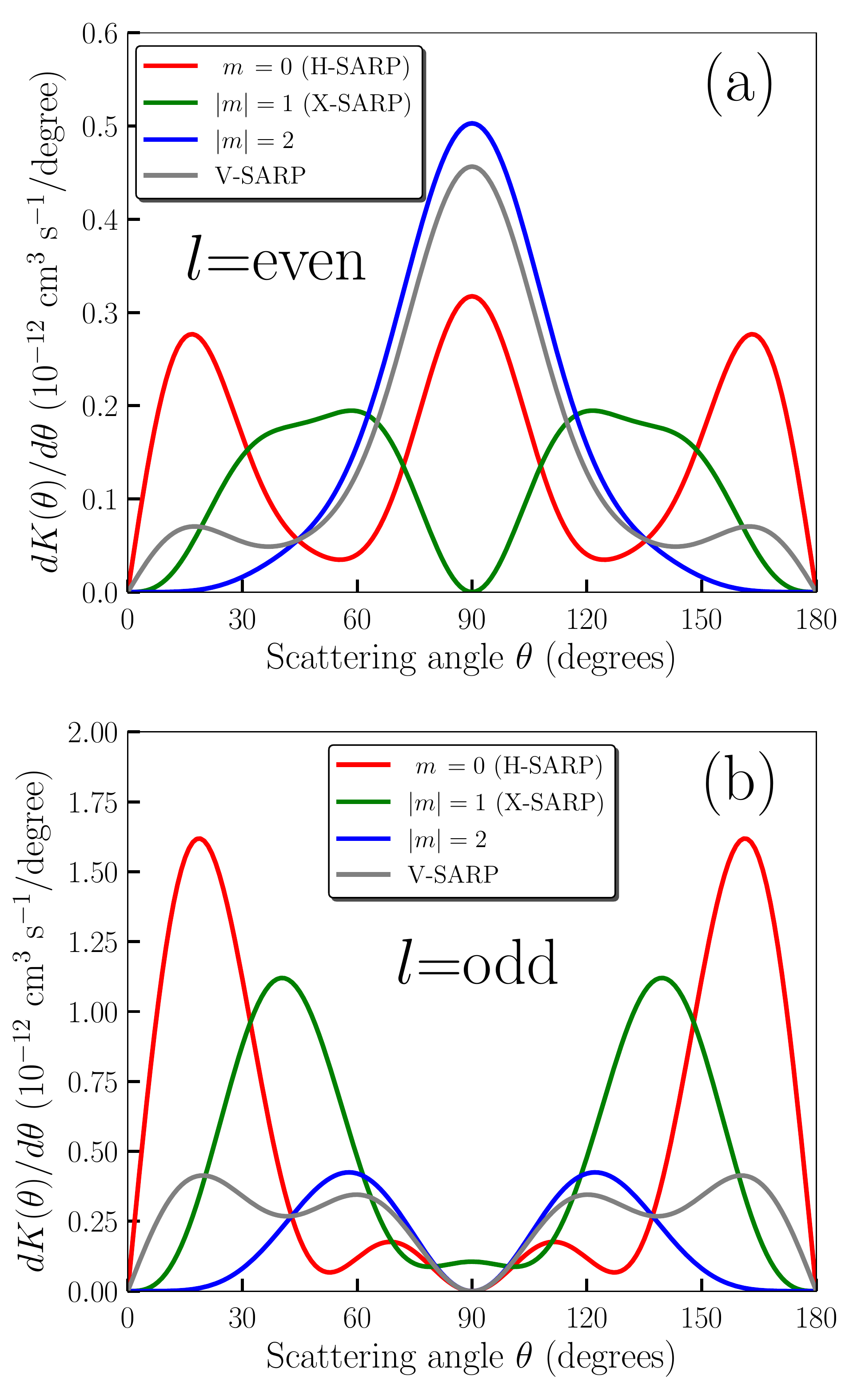}
\end{center}
\caption{Differential rate  constant 
for HD ($v=1,j=2 \to v'=1,j'=0$) by  collision with He as a function of the scattering angle. (a) even (b) odd partial waves. 
}
\label{fig:Drate_evenodd}
\end{figure}

Next, we consider angular distribution of inelastically scattered HD and contributions from even and odd partial waves.
To simulate the experimental angular distribution it is necessary to calculate the differential rate 
constant~\cite{2018_PRL_Croft} averaged over the relative velocity distribution of He and HD in the experiment. The angular distribution is given by~\cite{2018_PRL_Croft,SARP_HD-He} 
\begin{equation}
\label{eq:rate}
\frac{dK(\theta)}{d\theta} =
\int_{0}^\infty |v_r|P(v_r)\frac{d\sigma(\theta)}{d\theta}dv_r 
+ \int_{-\infty}^{0} |v_r|P(v_r)\frac{d\sigma(\pi-\theta)}{d\theta}dv_r.
\end{equation}
We note that the relative velocity distribution $P(v_r)$ reported by Perreault {\it et al}.~\cite{SARP_HD-He} is not symmetric about $v_r=0$ (see also Fig.~S3 in supplementary material).
The DCS ${d\sigma(\theta)}/{d\theta}$  depends on the relative velocity $v_r$ and the collision energy as given in \cref{eq:DCS}. In particular, near a resonance the DCS shows significant energy dependence (see also Fig.~S4--S7 in supplementary material).

\begin{figure}[b!]
\begin{center}
\includegraphics[scale=0.4]{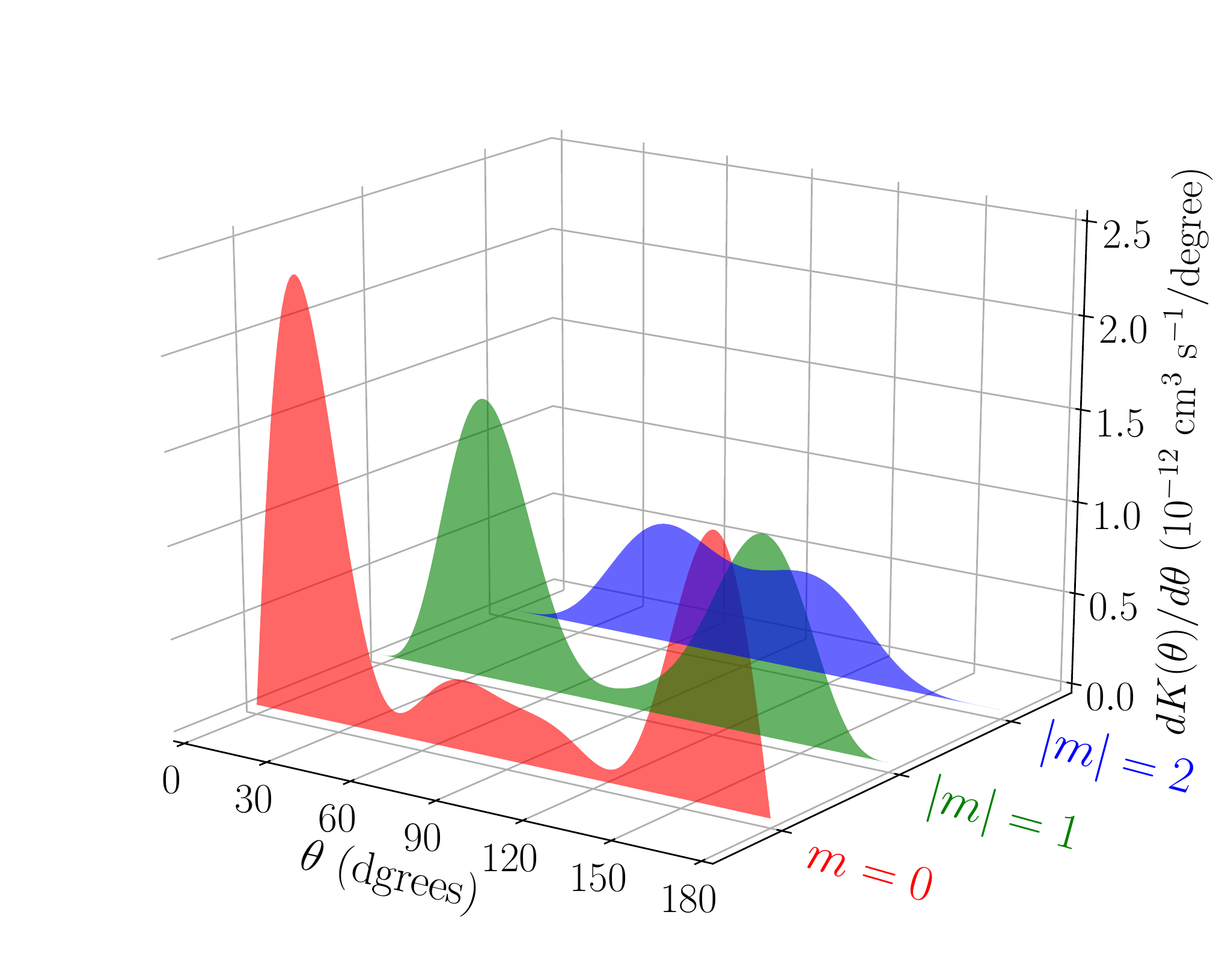}
\end{center}
\caption{Differential rate  constants 
for HD ($v=1,j=2 \to v'=1,j'=0$) by  collision with He as  functions of the scattering angle for different initial $|m|$-states of HD. Results include contributions from 
both even and odd partial waves.}
\label{fig:3Drate_evenodd}
\end{figure}

In \Cref{fig:Drate_evenodd}, we present the differential rate 
constants
for even and odd partial wave contributions derived from \cref{eq:rate}.
As discussed  by Perreault {\it et al}.,~\cite{SARP_HD-He} by choosing only even or odd partial waves, 
the symmetric angular distribution about $\theta=90^\circ$ is 
obtained regardless of the symmetry of the relative velocity distribution about $v_r=0$. 
The results exhibit distinctly different features for even and odd partial waves. 
The experimental results~\cite{SARP_HD-He} for H-SARP and V-SARP exhibit two prominent peaks centered around 25$^\circ$ and 165$^\circ$ with a shallow peak near 90$^\circ$. The shallow peak is more prominent for the V-SARP preparation. 
While the experimental results 
were attributed to
signatures of an $l=2$ resonance~\cite{SARP_HD-He}, the computed H-SARP and X-SARP
results for odd partial waves (primarily $l=1$) appears to be in better agreement with experiment. The computed V-SARP results show markedly different angular dependence compared to the experimental data and exhibit distinctly different features for even and odd $l$ values. In particular, the strong central peak seen for even $l$ is absent for odd $l$. The  $m$-resolved differential rate constants that include contributions from both even and odd partial waves are displayed in   \Cref{fig:3Drate_evenodd} as a function of  the scattering angle. The dominant contribution from $l=1$  is clearly reflected in the overall rate constant.  

\begin{figure}[b!]
\begin{center}
\includegraphics[scale=0.42]{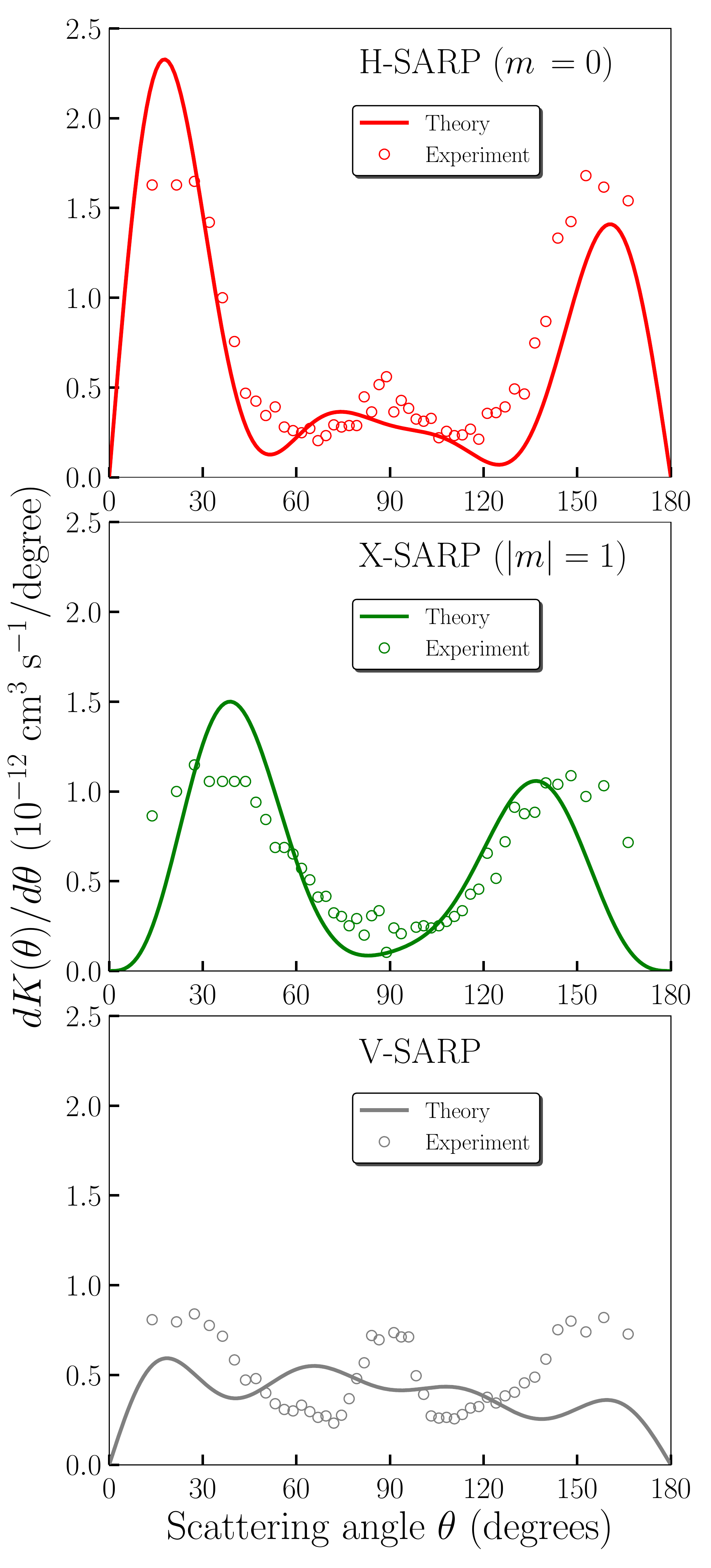}
\end{center}
\caption{Differential rate 
constant
for HD($v=1,j=2 \to v'=1,j'=0$) by collision with He as a function of the scattering angle. The calculated results 
are shown by solid curves and the experimental counts are shown by circles (by multiplying a constant factor of $4 \times10^{-15}$). 
(a) H-SARP ($m=0$) (b) X-SARP ($|m|=1$) and (c) V-SARP.   
}
\label{fig:Drate}
\end{figure}

In \Cref{fig:Drate} we compare our results that include contributions from both even and odd partial waves with the experimental angular distributions~\cite{SARP_HD-He}. 
For  the comparison, we omitted  contributions from DCS below 0.01 cm$^{-1}$ 
($|v_r| \sim 12$ m/s)
in evaluating the integrals in \cref{eq:rate} to limit contributions from purely $s$-wave scattering. However, this truncation has only a very small effect as higher or lower values of energy truncations yield very similar results  
(see Fig.~S8. in supplementary material).
Our theoretical results for H-SARP and X-SARP are still in close agreement with experiment even if  contributions  from both even and odd partial waves are included.
The main difference is that the symmetry about $\theta=90^\circ$ is now absent in the theoretical results. 
In the analysis of the experimental results~\cite{SARP_HD-He}, 
it was assumed that the relative velocity dependence of the DCS is negligible,  
leading to an additional assumption
that the  DCS is symmetric about $\theta=\pi/2$ to explain the observed symmetry of the angular distribution.  
However,  this assumption is not fully valid as DCS exhibits strong energy dependence, in particular in the vicinity of resonances, 
and  the experimental relative velocity distribution is not sufficiently narrow to capture just the peak of a single (partial wave) resonance.

In summary, we have carried out explicit quantum calculations of integral and differential cross sections for rotational quenching of state prepared HD ($v=1,\ j=2\  \to\ v'=1,\ j'=0$) in collisions with He atoms at relative collision energies below 10 K. Our results show a prominent $l=1$ shape resonance near 0.2 cm$^{-1}$ and a weak shoulder feature from an $l=2$ partial wave between 1-3 cm$^{-1}$. These two resonances control the stereodynamics of the $\ j=2\  \to \ j'=0$ rotational transition in HD.
We have also identified a universal trend for the stereodynamic preference in the low energy regime for the integral cross section 
and its features in the vicinity of an $l=1$ resonance. 
It is found that the interference effect due to cross terms in the square of modulus of the scattering amplitude plays a critical role in determining the stereodynamic preference at higher energies where higher-order partial wave contributes. This interference effect alters the  stereodynamic preference near an $l=2$ resonance. 
The angular distribution of the scattered HD molecule 
is found to be strongly influenced by the $l=1$ resonance. The calculated angular distributions for the H-SARP and X-SARP preparations are found to be in overall good agreement with experiment but do not depict a symmetric profile around $\theta=90^\circ$ due to the strong energy dependence of the scattering amplitude and the interference between even and odd partial wave contributions. The V-SARP results also show broad agreement with experiment except that the central peak
reflects contributions from both even and odd partial waves leading to a double peak structure. 
We believe, measurements that include the energy dependence of the angular distribution
will allow a more accurate comparison with experiment.

\section*{Supplementary material}
See the supplementary material associated with this article for
computational details of the scattering calculation, partial wave contributions in integral cross section, 
sensitivity to the potential energy surface, universal trend in the stereodynamic preference at low energies, 
relative velocity distribution of He and HD, energy dependence of the differential cross section,
and low energy cutoff effect to the differential rate constants.

\begin{acknowledgments}
We are grateful to Nandini Mukherjee, Richard Zare, James Croft, and Hua Guo  for helpful discussions. 
This work was supported by NSF grant No. PHY-1806334 (N.B.) and ARO MURI grant No. W911NF-19-1-0283 (N.B.).
\end{acknowledgments}

\section*{Data availability statement}
The data that support the findings of this study are available within the article and its supplementary material.

\bibliography{cite}



\clearpage
\onecolumngrid
\vspace{\columnsep}

\newcolumntype{Y}{>{\centering\arraybackslash}X}
\newcolumntype{Z}{>{\raggedleft\arraybackslash}X}

\setcounter{figure}{0}
\setcounter{equation}{0}
\setcounter{page}{1}

\renewcommand{\thepage}{S\arabic{page}}
\renewcommand{\thefigure}{S\arabic{figure}}
\renewcommand{\theequation}{S\arabic{equation}}

\onecolumngrid

\begin{center}
	\textbf{\huge Supplementary Material}
\end{center}

\begin{center}
\text{ \it \Large Stereodynamics of rotationally inelastic scattering in cold He+HD collisions} 
\end{center}

\begin{center}
\text{ \large Masato Morita and Naduvalath Balakrishnan} 
\end{center}

\text{}
\newline


\setcounter{section}{1}
\begin{center}
\textbf{ \large
\thesection{\label{sec:theory}.	Computation (Scattering calculation)}
}
\end{center}

The quantum mechanical scattering calculations are carried out with the MOLSCAT code (v.14)~\cite{MOLSCAT} in the framework of the time-independent scattering theory to address the rotational quenching of HD ($v=1$, $j=2$ $\to$ $v'=1$, $j'=0$) by collision with He in cold energy regime.
In Jacobi coordinates the Hamiltonian for the collision complex may be written ($\hbar=1$) as
\begin{equation}
{
\hat{\mathcal{H}} = 
- \frac{1}{2\mu R} \frac{d^2}{dR^2}R 
 + \frac{\hat{\bm{l}}^2}{2\mu R^2}
 + \hat{\mathcal{H}}_\text{HD}
 + V_\text{int}(R,r,\gamma)
},
\label{eq:Heff}
\end{equation}
where $\mu$ is the reduced mass of the He+HD complex ($\mu=1.721871434$ amu), $\hat{\bm{l}}$ is the orbital angular momentum operator for the relative motion of the collision partners, and $\hat{\mathcal{H}}_\text{HD}$ is the ro-vibrational Hamiltonian for the isolated HD molecule.  The interaction potential $V_\text{int}(R,r,\gamma)$ between He and HD is 
 generated from  the BSP3 potential energy surface (PES) for the HeH$_2$ system.~\cite{BSP3}

The time-independent Schr\"{o}dinger equation for the scattering is numerically solved based on the close-coupling (CC) scheme using the total angular momentum basis set for the angular variables in the space-fixed (SF) coordinate frame.~\cite{1960_Dalgarno_Arthurs} The total wavefunction is expanded as 
\begin{equation}
\label{eq:totjbasis}
\Psi^{JM} = \frac{1}{R} \sum^{}_{v,j,l}
F^{JM}_{vjl}(R) \frac{\chi^j_v(r)}{r}|JM(lj)>
\end{equation}
where $J$ is the quantum number for the total angular momentum $\hat{\bm{J}}$ of the collision complex, $M$ is the quantum number for the projection $\hat{{J}_\mathrm{z}}$ onto the SF z-axis, $F^{JM}_{vjl}(R)$ are the radial expansion coefficients in $R$, $\chi^j_v(r)$ denote vibrational eigenfunctions of HD with quantum numbers of $v$ and $j$,  and $|JM(lj)>$ denote total angular momentum basis functions for describing the angular degrees of freedom in the total wavefunction.

The coefficients $F^{JM}_{vjl}(R)$ satisfy the CC equations obtained by substituting \cref{eq:totjbasis} into the time-independent Schr\"{o}dinger equation $\hat{\mathcal{H}}\Psi=E\Psi$, where $E$ is the total energy:
\begin{equation}\label{eq:CC}
\Bigl{ [ }  \frac{1}{2\mu }\frac{d^2}{dR^2}\ -\frac{l(l+1)}{2\mu R^2} + E_C \Bigr{ ] } F^{J}_{vjl}(R) 
= \sum^{}_{v',j',l'} 
\int_0^{\infty} <J(lj)| \chi^{j}_{v}(r) 
  V_\text{int}(R,r,\gamma)
\chi^{j'}_{v'}(r) |J(l'j')> dr.
\end{equation}
Here, we omit $M$ since the CC equations are independent of $M$. The collision energy $E_C$ is given by $E_C=E-E_{v,j}$ where $E_{v,j}$ denote the rovibrational energies of HD obtained by solving the following eigenvalue  equation:
\begin{equation}
\Bigl{ [ } -\frac{1}{2\mu_\text{HD}} \frac{d^2}{dr^2} + \frac{j(j+1)}{2\mu_\text{HD} r^2} +V_\text{HD}(r) \Bigr {] } \chi^j_v(r) = E_{v,j} \chi^j_v(r),
\label{eq:mol}
\end{equation}
where $\mu_\text{HD}$ is the reduced mass of HD. For $V_\text{HD}(r)$ in the electronic ground state ($X ^{1}\Sigma^+_\text{g}$), we adopt a modified Schwenke's H$_\text{2}$ potential energy curve reported by Boothroyd {\it et al}.~\cite{HD_PEC_modify} Due to the conservation of the total angular momentum $J$ of the collision complex and the inversion parity $\epsilon_I=(-1)^{j+l}$, the CC equation are separated into  independent blocks with respect to the values of $J$ and $\epsilon_I (= \pm 1)$. In this paper, we solve CC equations with 8 $J$-blocks, $J_\text{min}=0$ to $J_\text{max}=7$.

To solve the CC equations we employ the modified log-derivative propagation method of Manolopoulos,~\cite{logD_Manolopoulos} and propagate the log-derivative matrix from 2 to 50 \AA\ with a step size of 0.05 \AA. The S-matrix is obtained by matching to the usual scattering boundary condition at each energy. The CC equations are solved at collision energies ranging from 10$^{-6}$ to 10 cm$^{-1}$.\\
\newpage

\setcounter{section}{2}
\begin{center}
\textbf{ \large
\thesection{\label{sec:Partial}.	Partial wave contributions}
}
\end{center}

\begin{figure}[h!]
\begin{center}
\includegraphics[scale=0.4]{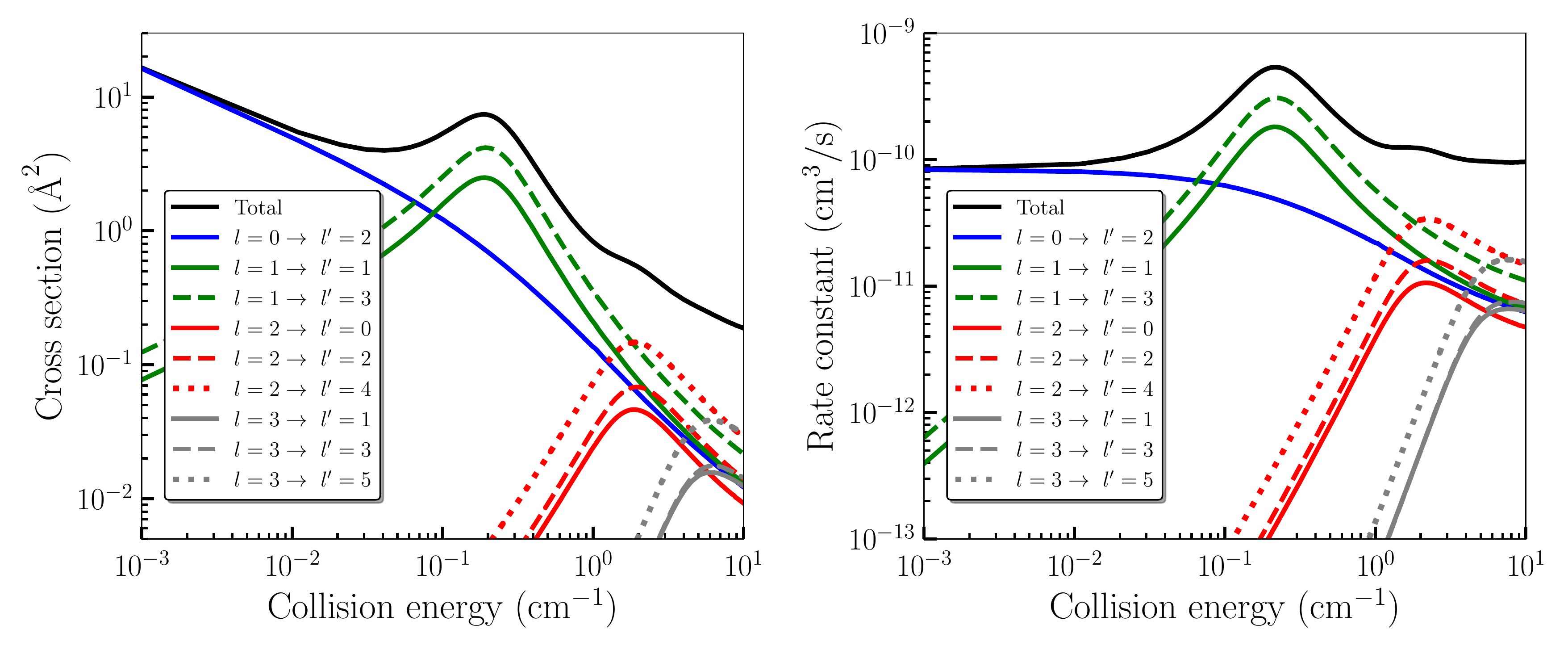}
\end{center}
\caption{
Left panel:  total and partial-wave resolved integral cross sections for HD ($v=1,j=2 \to v'=1,j'=0$) by collisions with He without preparation of the initial polarization of HD (isotropic). 
Right panel:  energy dependent-rate constants obtained by multiplying the integral cross section with the magnitude of relative collision velocity. \\ 
} 
\label{fig:ICS_partial}
\end{figure}

The left panel in \cref{fig:ICS_partial} shows the partial-wave resolved integral cross section (ICS) for HD ($v=1,j=2 \to v'=1,j'=0$) by collisions with He in which
the initial HD is non-polarized 
(isotropic). It is seen that the primary resonance at 0.2 cm$^{-1}$ originates from $l=1$ and $l'=1$ and 3 (green curves). The weak shoulder between 1-3 cm$^{-1}$ arises from $l=2$ and $l'=0,\,2$ and 4 (red curves). 
Importantly, at around the shoulder, the contributions from lower partial waves ($l=0$ and 1) are not necessarily negligible. For each $l$, the component associated with the  highest  possible $l'$ has the largest contribution but other lower $l'$ components are also important to describe the resultant total ICS, in particular, in the vicinity of the resonances. This indicates the difficulty of describing the total ICS/DCS using only a single outgoing partial wave $l'$ component even for an enhanced resonance like $l=1$.
In the right panel, we show the partial wave resolved energy dependent-rate constants obtained by multiplying the integral cross section with the magnitude of relative velocity for the collision. In the Wigner threshold regime where the $s$-wave dominates, the quenching cross section behaves as $1/\sqrt{E}$,
thus the rate constant becomes independent of the collision energy.
\\


\setcounter{section}{3}
\begin{center}
\textbf{ \large
\thesection{\label{sec:relative}.	Sensitivity to potential energy surface}}
\end{center}
\begin{figure}[hbt!]
\begin{center}
\includegraphics[scale=0.33]{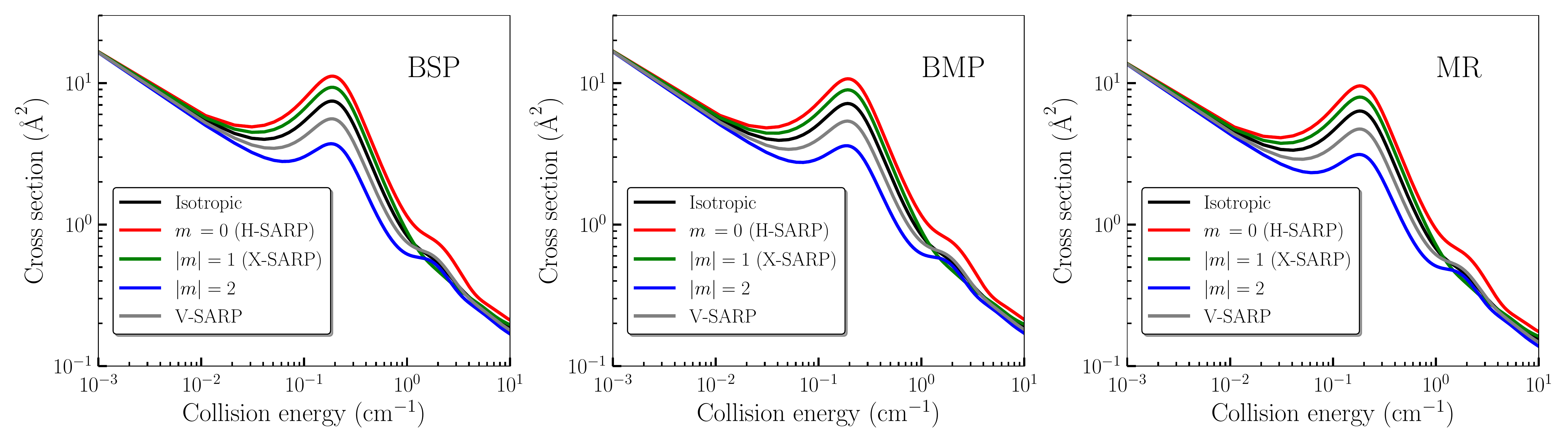}
\end{center}
\caption{
State-to-state integral cross sections for HD ($v = 1,j = 2 \to v'= 1,j' = 0$) by collisions with He using three different HeH$_2$ PESs: BSP~\cite{3D_PES} (left panel), BMP~\cite{HD_PEC_modify} (middle panel) and MR~\cite{HeH2_Muchnick_Russek} (right panel). Initial HD orientation ($m$) dependence is displayed with red ($m = 0$), green ($|m| = 1$), and blue ($|m| = 2$). Results for V-SARP preparation and isotropic collisions (non-polarized) are displayed in gray and black curves, respectively.
}
\label{fig:PESs}
\end{figure}

The scattering calculations are performed using the BSP3 PES which is the most accurate among available PESs for HeH$_2$. As discussed in the main text the BSP3 PES yields spectroscopic properties for He+H$_2$ and He+HD in excellent agreement with high precision experiments. To demonstrate that resonance features in the ICS are not sensitive to the choice of PES, we show the energy dependence of the ICS obtained with three other PESs~\cite{3D_PES,HD_PEC_modify,HeH2_Muchnick_Russek} which have been published earlier than BSP3 and widely used in previous calculations.  These are   
Bakr, Smith and Patkowski  (BSP) PES~\cite{3D_PES},  an improved version of Boothroyd, Martin and Peterson (BMP) PES \cite{HD_PEC_modify}, and a modified version of the Muchnick and Russek (MR) PES~\cite{HeH2_Muchnick_Russek}. The improvement and modification are given in
Bakr et al.~\cite{3D_PES}
We see that both BSP and BMP results are essentially identical to the BSP3 results in Fig. 1 (a) in the main text.
While the MR results (right panel) show an overall shift to the lower magnitudes, the behavior of the cross section including the resonance structure and initial orientation dependence is very similar to others. In all, the results presented in the main text with the BSP3 PES are very stable and robust with respect to small changes in the PES.
Thus, we can expect that the qualitative nature of the results presented here will not change even with any future improvement of the PES.\\
\\

\setcounter{section}{4}
\begin{center}
\textbf{ \large
\thesection{\label{sec:m-dependence} Universal trend in stereodynamic preference at low energies \\}}
\end{center}

As discussed in the main text, the initial $m$-dependence of the ICS in collisions between a structureless atom and a $^1\Sigma$ molecule is qualitatively determined by  the product of the following two 3-j symbols,
\begin{equation}
\begin{pmatrix} j & l & J\\ m & 0 & -m \end{pmatrix}   
\begin{pmatrix} j' & l' & J\\ m' & m-m' & -m \end{pmatrix}
\label{eq:CG1}
\end{equation}
In the present study, we consider the rotational quenching of HD ($j=2,m \to\ j'=0,m'=0$), and $J=l'$ due to $j'=0$, thus we need to investigate the property of 
\begin{equation}
\begin{pmatrix} 2 & l & J\\ m & 0 & -m \end{pmatrix}   
\begin{pmatrix} 0 & J & J\\ 0 & m & -m \end{pmatrix}.
\label{eq:CG2}
\end{equation}
In the s-wave limit, only one term associated with $l=0$ is effective in the scattering amplitude, i.e., $l=0$, $J=j=2$. The resulting value is independent of $m$:
\begin{equation}
\begin{pmatrix} 2 & 0 & 2\\ m & 0 & -m \end{pmatrix}   
\begin{pmatrix} 0 & 2 & 2\\ 0 & m & -m \end{pmatrix} 
= \frac{(-1)^{-m}}{\sqrt{5}} \times  \frac{(-1)^{-m}}{\sqrt{5}}   = \frac{1}{5}. 
\label{eq:CG3}
\end{equation}
The initial orientation ($m$) independence comes from purely algebraic character of the scattering amplitude, thus the $m$-independence of the ICS is universal for this class of atom-molecule systems in the low energy limit for the $j=2 \to j'=0$ quenching. 
We note that it is straightforward to show the $m$-independence for the quenching processes to $j'=0$ from arbitrary excited initial states $j$ with $l=0$.

For $j=2$, there are two terms associated with $l=1$ in the scattering amplitude: $J=l'=1$ and $J=l'=3$. For the former, the product of two 3-j symbol becomes~\cite{Mathematica}
\begin{equation}
\begin{pmatrix} 2 & 1 & 1\\ m & 0 & -m \end{pmatrix}   
\begin{pmatrix} 0 & 1 & 1\\ 0 & m & -m \end{pmatrix} 
= \frac{(-1)^{-m} \sqrt{4-m^2} }{\sqrt{30}} \times  \frac{(-1)^{1-m}}{\sqrt{3}}   = \frac{-\sqrt{4-m^2} }{\sqrt{90}}.
\label{eq:CG1p}
\end{equation}
Since the square of the modulus of the scattering amplitude determines the magnitude of the cross section, the ratio of the ICS is 4:3 for $m=0$ vs $|m|=1$, and $|m|=2$ results in 0.  For the $J=l'=3$ case,~\cite{Mathematica}
\begin{equation}
\begin{pmatrix} 2 & 1 & 3\\ m & 0 & -m \end{pmatrix}   
\begin{pmatrix} 0 & 3 & 3\\ 0 & m & -m \end{pmatrix} 
= \frac{(-1)^{1+m} \sqrt{9-m^2} }{\sqrt{105}} \times  \frac{(-1)^{1-m}}{\sqrt{7}}   = \frac{\sqrt{9-m^2} }{\sqrt{735}}.
\label{eq:CG2p}
\end{equation}
The ratio of the ICSs becomes 9:8:5 for $m=0$, $|m|=1$ and $|m|=2$, respectively.  Importantly, the stereodynamic preference is similar for both $J=1$ and 3, and the  ICS displays the following trend in low energy regime:
$m=0$ (H-SARP) $>$ $|m|=1$ (X-SARP)  $>$ (V-SARP) $>$ $|m|=2$. 
However, the relative contribution between $J=1$ and 3 is system and energy dependent because it is determined by the values of the respective T-matrix elements. \\ 
\\
\newpage

\setcounter{section}{5}
\begin{center}
\textbf{ \large
\thesection{\label{sec:relative}.	Relative velocity distribution}}
\end{center}

\begin{figure}[hb!]
\begin{center}
\includegraphics[scale=0.38]{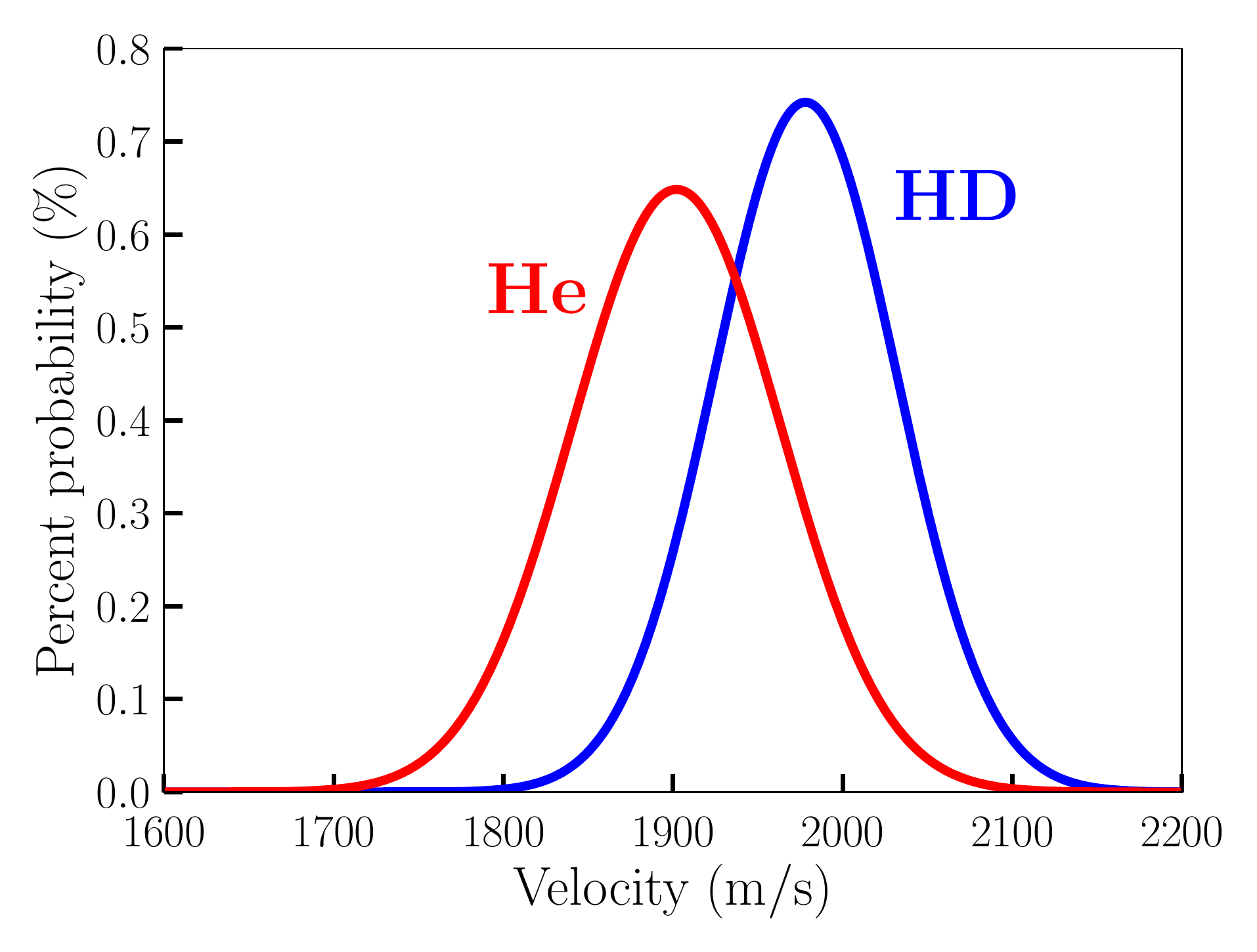}
\includegraphics[scale=0.38]{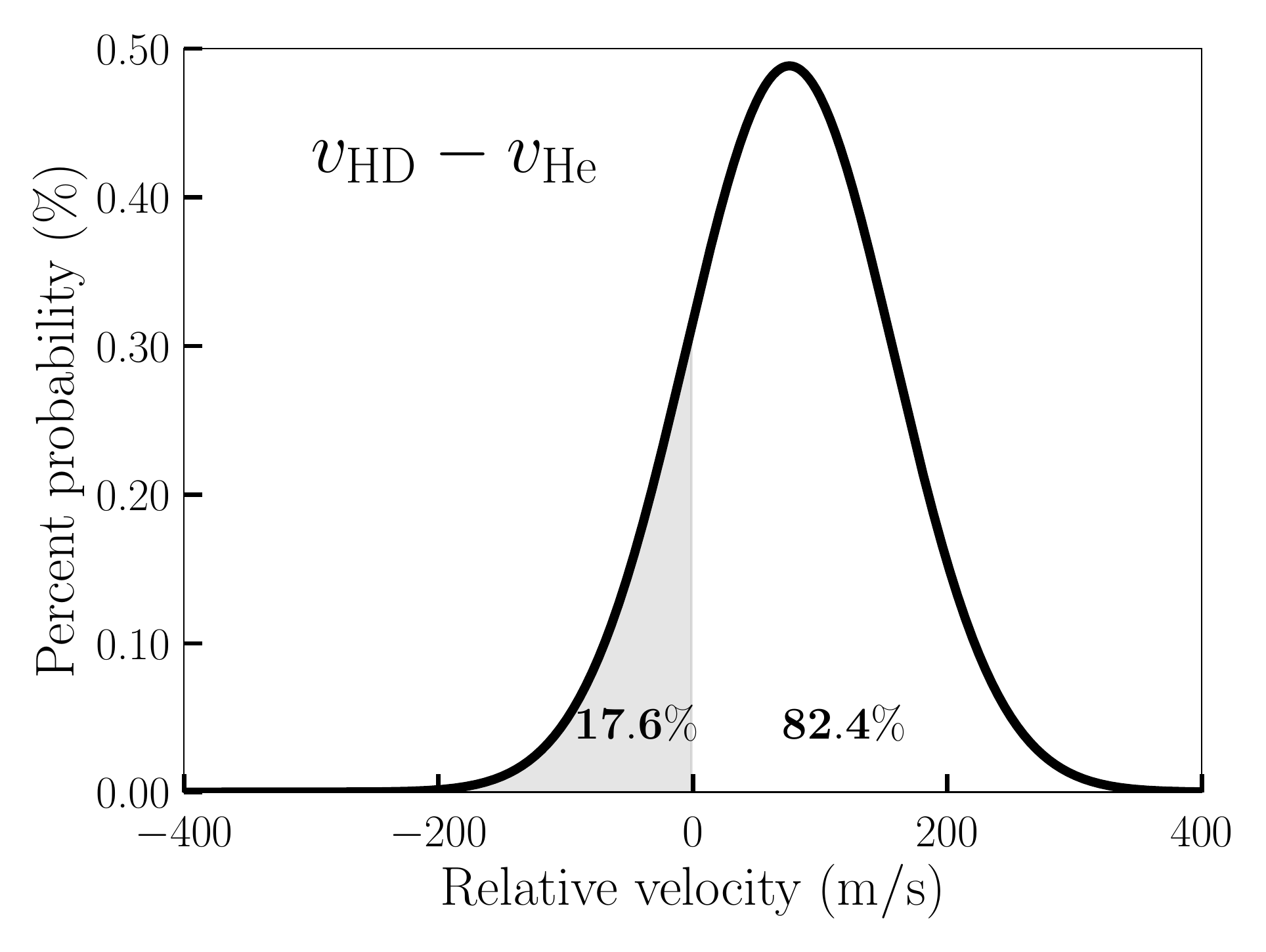}
\end{center}
\caption{Left panel: velocity distributions for HD and He. Right panel: relative velocity distribution between HD and He. 
}
\label{fig:velocity_dist}
\end{figure}

To directly compare with the experimentally observed angular distribution of inelastically scattered HD reported by Perreault {\it et al}.,~\cite{SARP_HD-He} we average our computed DCS with the relative velocity distribution obtained from the velocity distributions reported by Perreault {\it et al}.
The velocities $v$ ($\text{m/s}$)  of HD and He are expressed as Gaussian functions in their supplementary material as 
\begin{equation}
\label{eq:velocity_dist}
P(v_\text{HD};\mu_\text{HD},\sigma^2_\text{HD}) =\frac{1}{\sqrt{2\pi}\sigma_\text{HD}}\text{exp}\{-\frac{(v_\text{HD}-\mu_\text{HD})^2}{2\sigma_\text{HD}^2}\}\ \ \ \text{and}\ \ \   
P(v_\text{He};\mu_\text{He},\sigma^2_\text{He}) =\frac{1}{\sqrt{2\pi}\sigma_\text{He}}\text{exp}\{-\frac{(v_\text{He}-\mu_\text{He})^2}{2\sigma_\text{He}^2}\}, 
\end{equation}
where $\mu_\text{HD}=1978$ and $\sigma_\text{HD}=76/\sqrt{2}$ for HD and $\mu_\text{He}=1902$ and $\sigma_\text{He}=87/\sqrt{2}$, respectively. The parameters $\mu, ~\sigma$, and $v$ are in units of m/s. While the Gaussian normalization factor of $1/(\sqrt{2\pi}\sigma)$ is omitted in the supplementary material of the experimental paper, the factor is explicitly included in \cref{eq:velocity_dist} for completeness (\cref{fig:velocity_dist} left panel).  

The distribution of the relative (1D) velocity $v_r$ of HD with respect to He ($v_r=v_\text{HD}-v_\text{He}$) is obtained by the convolution of the above velocity distributions. 
The distribution of $v_r$ given by the convolution of the Gaussians is a Gaussian as a consequence of the reproductive property of normal distribution:
\begin{equation}
\begin{split}
\label{eq:speed_dist}
P(v_r) =
\int_{-\infty}^\infty P(v_\text{HD};\mu_\text{HD},\sigma^2_\text{HD})P(v_\text{HD}-v_r;\mu_\text{He},\sigma^2_\text{He}) dv_\text{HD}  
&= {\frac{1}{\sqrt{2\pi (\sigma_\text{HD}^2+\sigma_\text{He}^2)}  }\text{exp}\{ -\frac{(v_r-(\mu_\text{HD}-\mu_\text{He}))^2}{2(\sigma_\text{HD}^2+\sigma_\text{He}^2)}} \} \\
&= P(v_r;\mu_\text{HD}-\mu_\text{He},\sigma_\text{HD}^2+\sigma_\text{He}^2).
\end{split}
\end{equation}

The resulting experimentally derived
relative velocity distribution~\cite{SARP_HD-He} is shown in the right panel of \cref{fig:velocity_dist}. As described in the supplementary materials of Ref.~[\onlinecite{SARP_HD-He}], the distribution is asymmetric about zero and the contributions of positive and negative relative velocities are $82.4\%$ and $17.6\%$, respectively. This result is consistent with the velocity distributions in the left panel since the distribution for HD expands in a slightly higher velocity region compared to He. \\
\\

As shown in the main text, we can obtain the angular distribution of inelastically scattered HD in the form of differential rate constant
($dK/d\theta$) from the relative velocity distribution $P(v_r)$ and the DCS ($d\sigma/d\theta$) as follows \cite{2018_PRL_Croft};
\begin{equation}
\label{eq:rate}
\frac{dK(\theta)}{d\theta} =
\int_{0}^\infty |v_r|P(v_r)\frac{d\sigma(\theta)}{d\theta}dv_r 
+ \int_{-\infty}^{0} |v_r|P(v_r)\frac{d\sigma(\pi-\theta)}{d\theta}dv_r. \\
\end{equation}
\\
\newpage


\setcounter{section}{5}
\begin{center}
\textbf{ \large
\thesection{\label{sec:DCS}.	Energy dependence of differential cross section}}
\end{center}

\begin{figure}[b!]
\includegraphics[scale=0.245]{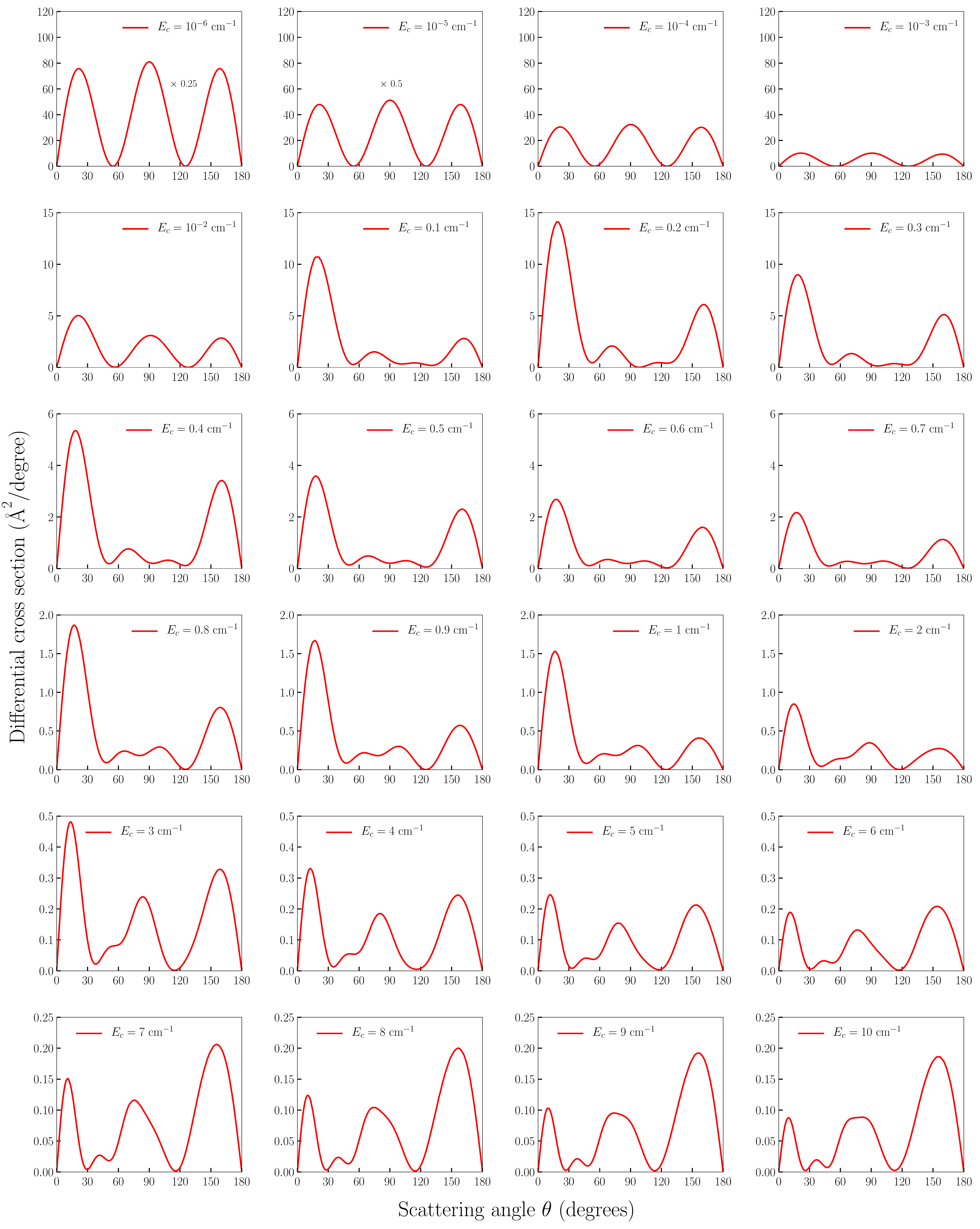}
\caption{Differential cross section for rotational quenching of oriented HD ($v=1, j=2, m=0$\ : HSARP) to ($v'=1, j'=0$) by collision with He at selected collision energies. Note that the range of the vertical axis depends on the collision energy. At $E_c=10^{-6}\ \text{and}\ 10^{-5}$ cm$^{-1}$, the cross section is scaled by factors of 0.25 and 0.5, respectively. \\
}
\label{fig:DCS_k0}
\end{figure}

\begin{figure}[b!]
\begin{center}
\includegraphics[scale=0.245]{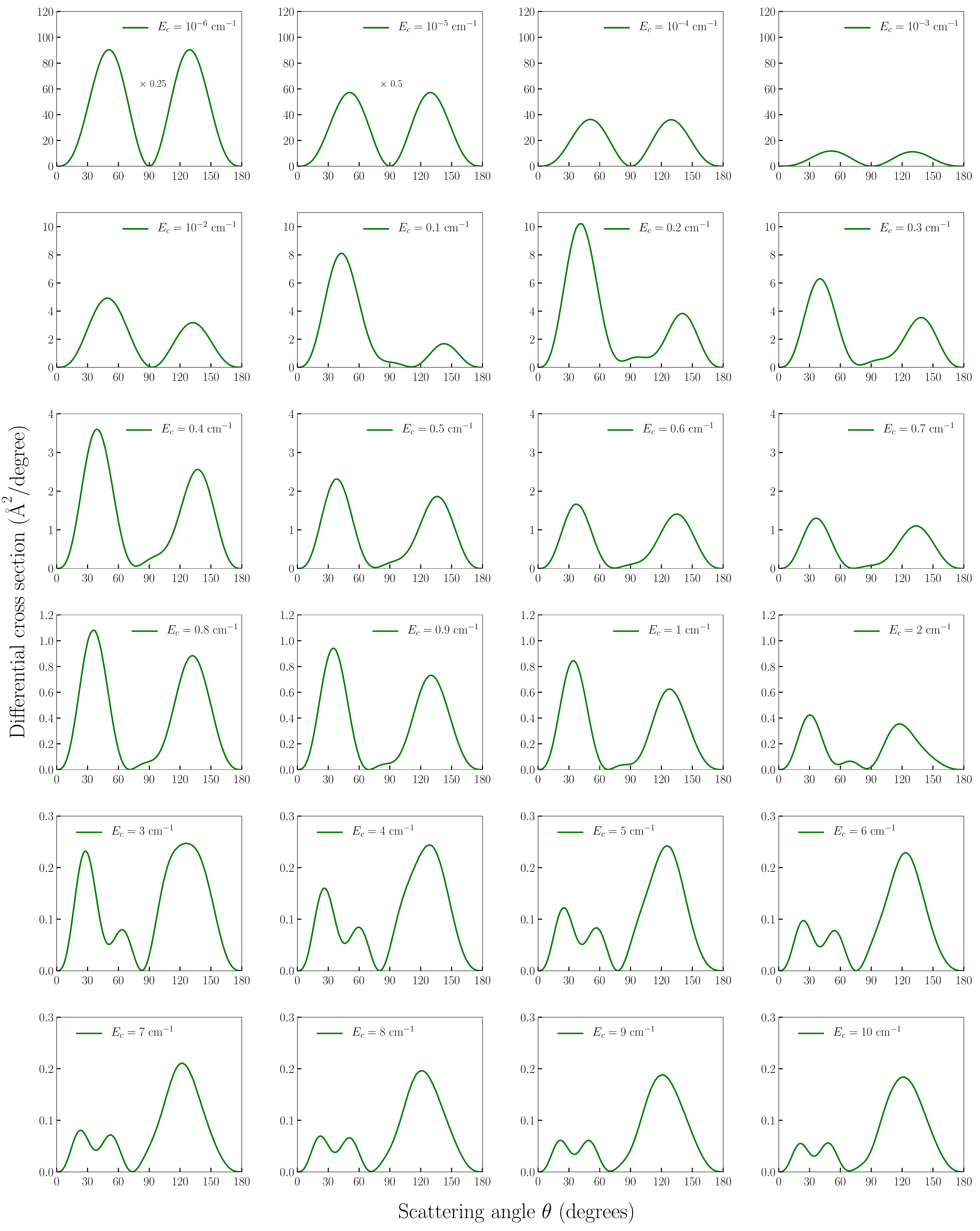}
\end{center}
\caption{Differential cross section for rotational quenching of oriented HD ($v=1, j=2, m=1\ \text{or}\ -1$  \ \text{or}\  X-SARP($m=\pm1$) to ($v'=1, j'=0$) by  collision with He at selected collision energies. Note that the range of the vertical axis depends on the collision energy. At $E_c=10^{-6}\ \text{and}\ 10^{-5}$ cm$^{-1}$, the cross section is scaled by factors of 0.25 and 0.5, respectively. \\
}
\label{fig:DCS_k1}
\end{figure}

\begin{figure}[b!]
\begin{center}
\includegraphics[scale=0.245]{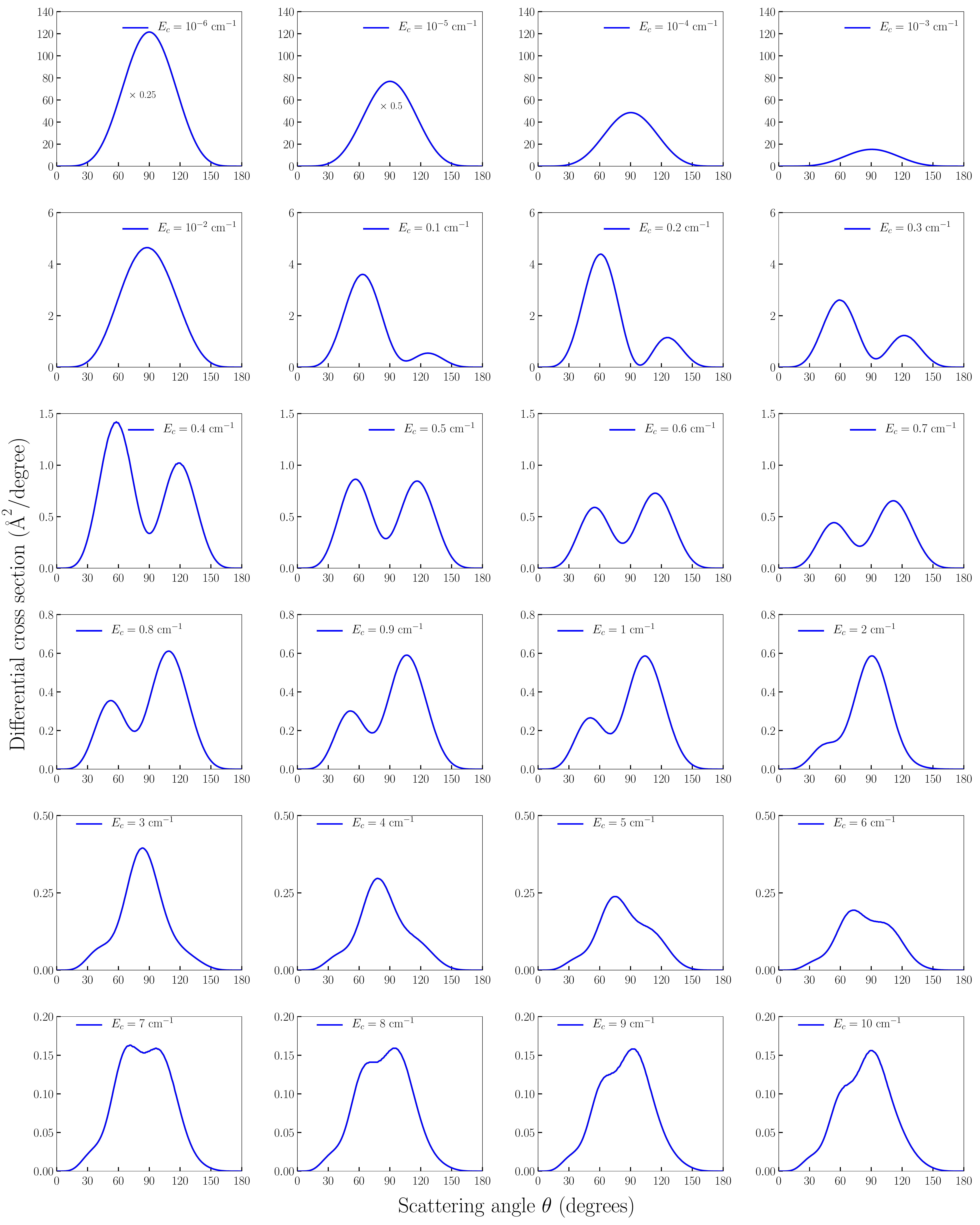}
\end{center}
\caption{Differential cross section for rotational quenching of oriented HD ($v=1, j=2, m=2\ \text{or}\ -2$) to ($v'=1, j'=0$) by collision with He at selected  collision energies. Note that the range of the vertical axis depends on the collision energy. At $E_c=10^{-6}\ \text{and}\ 10^{-5}$ cm$^{-1}$, the cross section is scaled by factors of 0.25 and 0.5, respectively. \\
}
\label{fig:DCS_k2}
\end{figure}

\begin{figure}[b!]
\begin{center}
\includegraphics[scale=0.245]{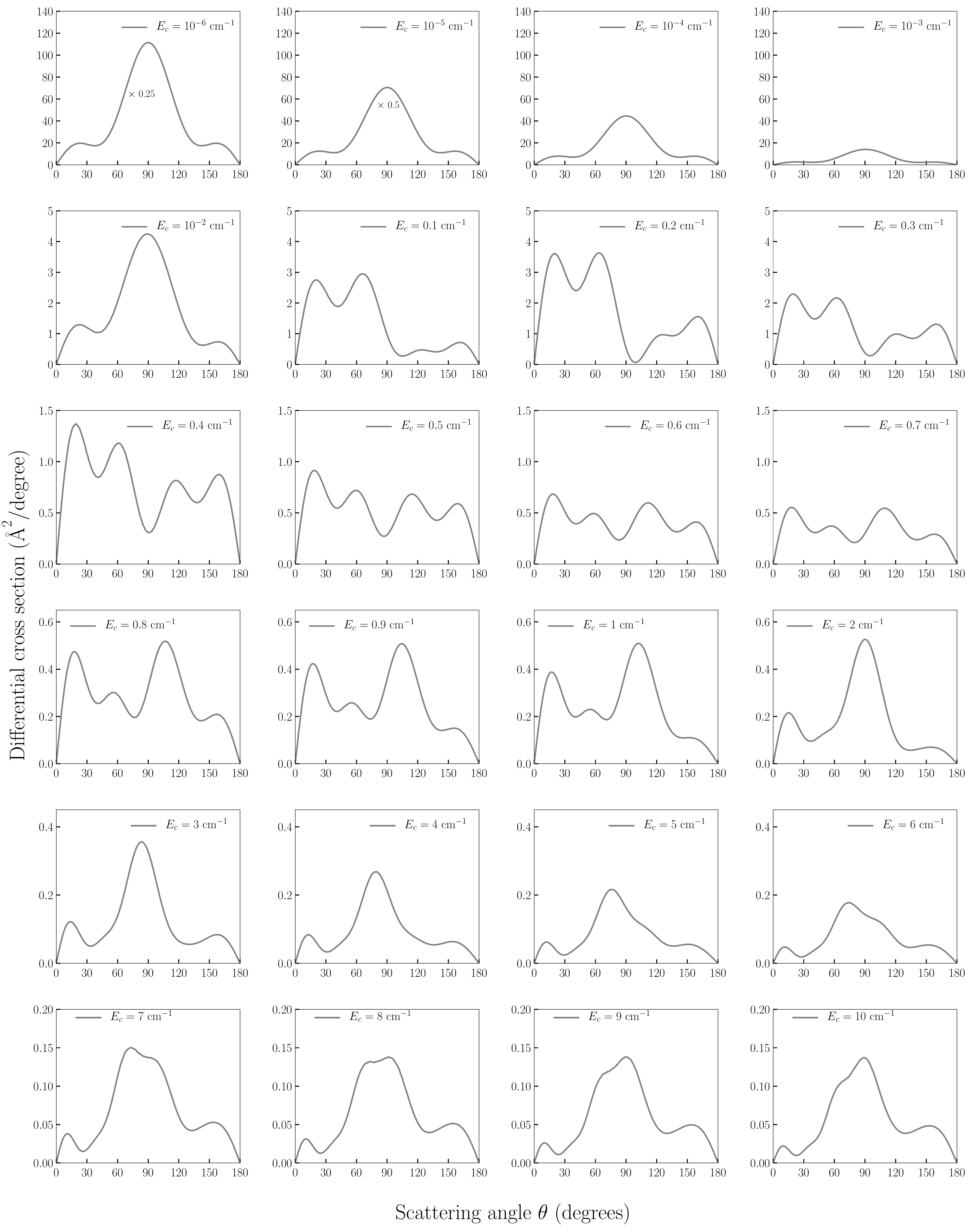}
\end{center}
\caption{Differential cross section for rotational quenching of aligned HD with V-SARP preparation, the HD bond axis aligned vertical to the initial relative velocity, ($v=1, j=2$ : V-SARP) to ($v'=1, j'=0$) by collision with He at selected collision energies. Note that the range of the vertical axis depends on the collision energy. At $E_c=10^{-6}\ \text{and}\ 10^{-5}$ cm$^{-1}$, the cross section is scaled by factors of 0.25 and 0.5, respectively.
}
\label{fig:DCS_V-SARP}
\end{figure}

The differential cross section (DCS)  ($d\sigma/d\theta$) in \cref{eq:rate} has an energy dependence due to the energy dependence of the S-matrix in the scattering amplitudes
and the $1/k^2$ factor, where $k$ is the magnitude of the wave vector in the incident channel.
The  S-matrix elements exhibit strong energy dependence in the vicinity of scattering resonances.
In \cref{fig:DCS_k0,fig:DCS_k1,fig:DCS_k2,fig:DCS_V-SARP} we show the energy dependence of the DCS in the relevant energy region with various initial HD orientations (\underline{figures start from the page S6}). Indeed, we see a significant energy dependence of the DCS around the resonances. \\
\\

\setcounter{section}{6}
\begin{center}
\textbf{ \large
\thesection{\label{sec:Emin}.   Low energy cutoff effect on the differential rate constant}}
\end{center}

\begin{figure}[b!]
\begin{center}
\includegraphics[scale=0.33]{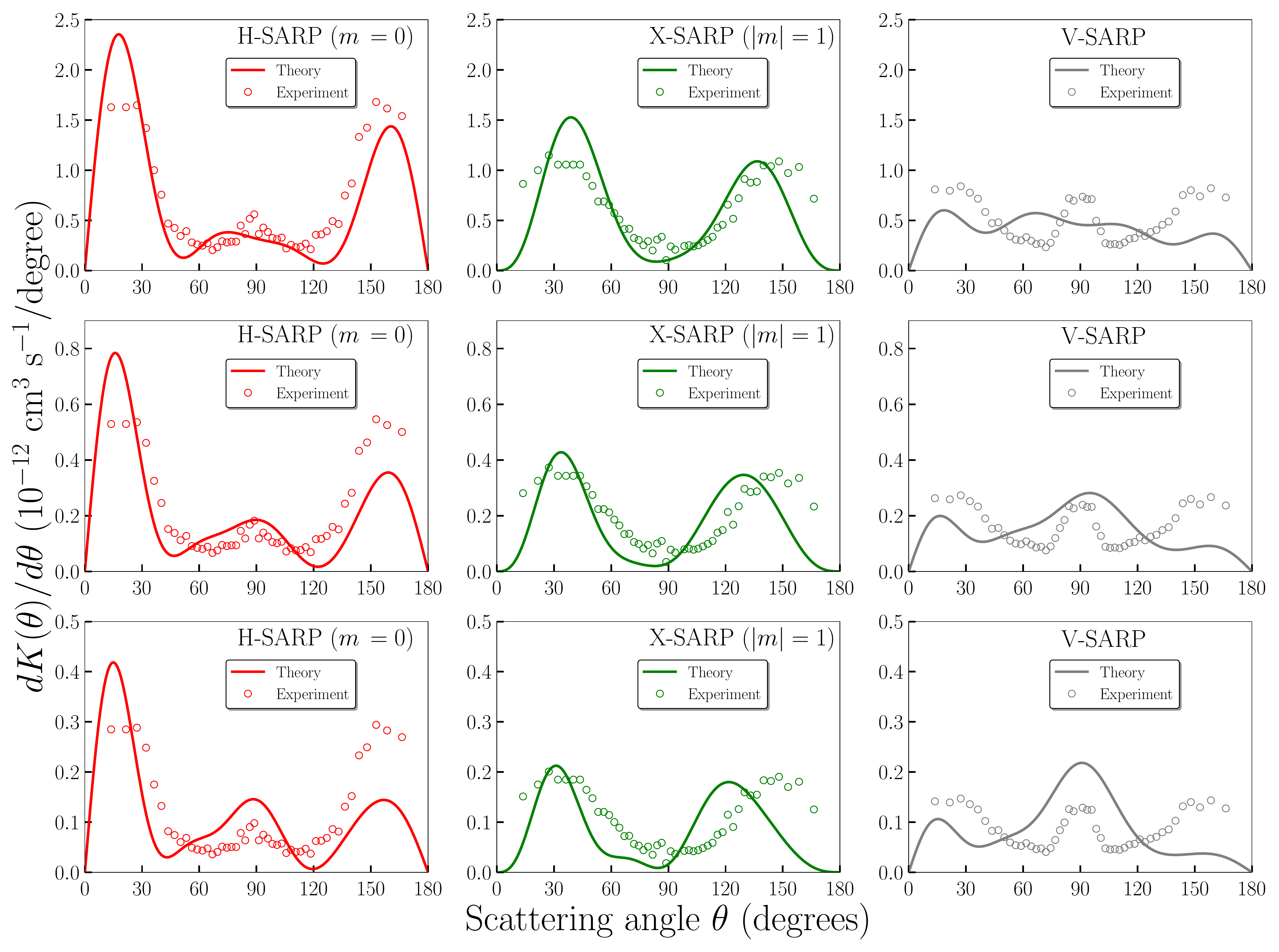}
\end{center}
\caption{
Differential rate constant for HD($v=1,j=2 \to v'=1,j'=0$) by collision with He as a function of the scattering angle. 
The calculated results are displayed with solid curves and the experimental counts are shown by circles (a) H-SARP ($m=0$) (b) X-SARP ($|m|=1$) and (c) V-SARP.   \\
Upper panels:  without a cutoff for the lower energy limit for the integrals in \cref{eq:rate} (Experimental results are multiplied by a constant factor of $4 \times10^{-15}$). \\
Middle panels: with a lower energy cutoff of 0.5 cm$^{-1}$ for the integrals in \cref{eq:rate} (Experimental results are  multiplied by a constant factor of $1.3 \times10^{-15}$). \\
Bottom panels:  with a lower energy cutoff of 1.0 cm$^{-1}$ for  the integrals in \cref{eq:rate} (Experimental results are multiplied by a constant factor of $0.7 \times10^{-15}$).
}
\label{fig:Emin}
\end{figure}

In Fig.~3 of the main text, we compared the velocity averaged angular distributions with the experimental results. 
In the velocity average we restricted the lower  limit of $|v_r|$ to $\sim$ 12 m/s ($E_\text{min}=0.01$ cm$^{-1}$) in the integrals in \cref{eq:rate}. 
To demonstrate that this  cutoff for the lower energy limit (lower/upper velocity limits for the integrals in \cref{eq:rate}) has no significant effect on the final results,  we show  results with no restriction on the lower limit ($E_\text{min}\sim 0$ cm$^{-1}$) and with $E_\text{min}=0.5$ cm$^{-1}$ ({$|v_r|$}$\sim$ 83 m/s) and 1.0 cm$^{-1}$ ($|v_r|$$\sim$ 118 m/s) in \cref{fig:Emin} (page S10).  
For H-SARP and X-SARP, the results without energy cutoff exhibit better overall agreement with experiments compared to those with cutoff values of 0.5 and 1.0 cm$^{-1}$.
On the other hand, for V-SARP,  the peak at $\theta=90^\circ$ is more accurately described with an energy cutoff of 0.5 cm$^{-1}$ or 1.0 cm$^{-1}$.
\\
\\
\\
\\

\end{document}